# Tunneling induced dark states and controllable fluorescence spectrum in quantum-dot molecules

Si-Cong Tian[1,†], Ren-Gang Wan[2], Cun-Zhu Tong[1,*], Yong-Qiang Ning[1], Li-Jun Wang[1]

[1] *State Key laboratory of Luminescence and Applications, Changchun Institute of Optics, Fine Mechanics and Physics, Chinese Academy of Sciences, Changchun 130033, China*

[2] *State Key Laboratory of Transient Optics and Photonics, Xi'an Institute of Optics and Precision Mechanics, Chinese Academy of Sciences, Xi'an, 710119, China*

Corresponding author: [†] *tiansicong@ciomp.ac.cn*

[*] *tongcz@ciomp.ac.cn*

**ABSTRACT**

We theoretically investigate the spectrum of the fluorescence from triple quantum-dot molecules and demonstrate that it is possible to use tunneling to induce dark states. Unlike the atomic system, in quantum-dot molecules we can use tunneling to create the dark states and control fluorescence emission, requiring no coupling lasers. And interesting features such as quenching and narrowing of the fluorescence can be obtained. We also explain the spectrum with the transition properties of the dressed states generated by the coupling of the laser and the two tunneling. The quenching of the fluorescence is due to the tunneling induced dark states, while the narrowing of the central peak is due to the slow decay rate of the dressed levels.

# I. INTRODUCTION

In the past few decades, quantum coherence was shown to be essential for many applications of multilevel quantum systems driven by laser fields and have resulted in a lot of interesting phenomena. One example is the dark resonances or coherent population trapping [1,2]. It is a basis for the effects such as electromagnetically induced transparency (EIT) [3,4], stimulated Raman adiabatic passage (STIRAP) [5,6], subluminal and superluminal light propagation [7,8], resonant enhancement of optical nonlinearity [9,10], spontaneous emission control [11,12] and lasing without inversion (LWI) [13,14].

The above studies are based on atomic systems using the coherent control lasers. In semiconductor quantum dots, confined electrons and holes exhibit atom-like properties, encouraging us to extend above studies to a solid system. Single quantum dots coherently driven by strong electromagnetic fields have been used to investigate quantum interference phenomena such as Autler-Townes splitting (ATS) and Mollow triplets [15,16], EIT [17], and resonance fluorescence [18,19]. Moreover, two or more quantum dots coupled by tunneling can form quantum dot molecules (QDMs). In such molecule, the tunneling of electrons or holes can be controlled by an external electric field and create a multilevel structure of excitonic states. Many works have been carried out about double quantum dot molecules (DQDs), such as optical spectroscopy [20], EIT and slow light [21], excitonic entanglement [22], protection of quantum states [23], controlled rotation of exciton qubits [24] and exciton-spin memory [25].

For optical spectroscopy is a powerful tool for probing and manipulating QDs, in this paper, we investigate the fluorescence spectrum from triple quantum dot molecules (TQDs) under the tunneling coupling, using the methods of atomic physics. Such molecules have been achieved in much experimental progress [26-29]. The fluorescence spectrum can acquire quenching, which is resulted from the dark states induced by the tunneling between the dots. Unlike the atomic system, in QDMs, the dark states is induced by the tunneling requiring no extra laser fields. And the fluorescence spectrum can also acquire narrowing of the central peak due to the slow decay rate of the dressed state generated by the tunneling and the laser field.

The paper is organized as follows: in Sec. II, we introduce the model and the basic equations. In Section III we describe the numerical results and explain the corresponding features. Section IV is the conclusions.

## II. TRIPLE QUANTUM DOT SYSTEM

We show the schematic of the band structure and level configuration of a TQD system in Fig. 1. From Fig. 1(a), without a gate voltage, the conduction-band electron levels are out of resonance and the electron tunneling between the QDs is very weak. While, Fig. 1(b) shows the case with a gate voltage, and the conduction-band electron levels come close to resonance and the electron tunneling between the QDs is greatly enhanced. And in the latter case the hole tunneling can be neglected because of the more off-resonant valence-band energy levels. Thus we can give the schematic of the level configuration of a TQD system, as shown in Fig. 1(c). Without the excitation of the laser, no excitons are inside all QDs, which corresponds to state $|0\rangle$. When a laser field is applied, a direct exciton is created inside the QD 1, condition represented by the state $|1\rangle$. Under the tunneling couplings, the electron can tunnel from QD 1 to the QD 2, and from QD 2 to QD 3. And we denote these indirect the excitons as state $|2\rangle$ and state $|3\rangle$. In such system, we can controll the tunnel barrier by placing a gate electrode between the neighboring dots.

In the interaction picture and the rotating wave and dipole approximations, the Hamiltonian of this system is (we use units such that $\hbar = 1$)

$$H = \sum_{j=0}^{3} E_j |j\rangle\langle j| + [(\Omega_c e^{-i\omega_c t}|0\rangle\langle 1| + T_1|2\rangle\langle 1| + T_2|3\rangle\langle 2|) + \text{H.c.}], \tag{1}$$

where $E_j = \hbar\omega_j$ is the energy of state $|j\rangle$, $\omega_c$ is the laser frequency, $\Omega_c = \boldsymbol{\mu}_{01} \cdot \mathbf{e} \cdot E$ is the Rabi frequency of the transition $|0\rangle \to |1\rangle$, with $\boldsymbol{\mu}_{01}$ being the associated dipole transition-matrix element, $\mathbf{e}$ the polarization vector and $E$ the electric-field amplitude of the laser pulse. And $T_1$ and $T_2$ are the tunneling coupling.

From the Liouville equation, we obtain the following equations for the density-matrix elements:

$$\dot{\rho}_{01} = i[\Omega_c(\rho_{11} - \rho_{00}) - T_1\rho_{02}] + (i\delta_1 - \Gamma_{10})\rho_{01}, \tag{2a}$$

$$\dot{\rho}_{02} = i(\Omega_c\rho_{12} - T_1\rho_{01} - T_2\rho_{03}) + \frac{1}{2}[i(\delta_1 + \delta_2) - \Gamma_{20}]\rho_{02}, \tag{2b}$$

$$\dot{\rho}_{03} = i(\Omega_c\rho_{13} - T_2\rho_{02}) + \frac{1}{2}[i(\delta_1 + \delta_3) - \Gamma_{30}]\rho_{03}, \tag{2c}$$

$$\dot{\rho}_{11} = i[\Omega_c(\rho_{01} - \rho_{10}) + T_1(\rho_{21} - \rho_{12})] - \Gamma_{10}\rho_{11}, \tag{2d}$$

$$\dot{\rho}_{12} = i[T_1(\rho_{22} - \rho_{11}) + i\Omega_c\rho_{02} - T_2\rho_{13}] + \frac{1}{2}[i(\delta_1 - \delta_2) - \Gamma_{10} - \Gamma_{20}]\rho_{12}, \tag{2e}$$

$$\dot{\rho}_{13} = i[\Omega_c\rho_{03} + T_1\rho_{23} - T_2\rho_{12}] + \frac{1}{2}[i(\delta_1 - \delta_3) - \Gamma_{10} - \Gamma_{30}]\rho_{13}, \tag{2f}$$

$$\dot{\rho}_{22} = iT_1(\rho_{12} - \rho_{21}) + iT_2(\rho_{32} - \rho_{23}) - \Gamma_{20}\rho_{22}, \tag{2g}$$

$$\dot{\rho}_{23} = i[T_2(\rho_{33} - \rho_{22}) + T_1\rho_{13}] + \frac{1}{2}[i(\delta_2 - \delta_3) - \Gamma_{20} - \Gamma_{30}]\rho_{23}, \tag{2h}$$

$$\dot{\rho}_{33} = iT_2(\rho_{23} - \rho_{32}) - \Gamma_{30}\rho_{33}, \tag{2i}$$

$$\dot{\rho}_{ij} = -\dot{\rho}_{ji}^*, \tag{2j}$$

$$\rho_{00} + \rho_{11} + \rho_{22} + \rho_{33} = 1. \tag{2k}$$

Here the detunings are defined as $\delta_1 = (\omega_{10} - \omega_c)$, $\delta_2 = \delta_1 + 2\omega_{12}$ and $\delta_3 = \delta_1 + 2\omega_{12} + 2\omega_{23}$, with $\omega_{mn}$ the transition frequency between $|m\rangle$ and $|n\rangle$ states. And $\Gamma_{10}$, $\Gamma_{20}$ and $\Gamma_{30}$ are the radiative decay rate of populations from $|1\rangle \to |0\rangle$, $|2\rangle \to |0\rangle$ and $|3\rangle \to |0\rangle$. For simplicity, all the parameters are scaled by the decay rate $\Gamma_{10}$.

In order to theoretically demonstrate the dark states induced by the tunneling of the QD system, we evaluate the fluorescence spectrum. Following the common method used in Ref. [30], we calculate the steady state fluorescence spectrum. As is well known, the fluorescence spectrum is proportional to the Fourier transform of the steady state correlation function $\lim_{t \to \infty} \langle E^{(-)}(r, \tau + t) \cdot E^{(+)}(r, t) \rangle$ [31], where $E^{(\pm)}(r, t)$ are the positive and negative frequency parts of the radiation field in the far zone, which consists of a free field operator, and a source-field operator that is proportional to the atomic polarization operator. Thus the incoherent fluorescence spectrum $S(\omega)$ can be expressed in terms of the atomic correlation function

$$S(\omega) = \operatorname{Re} \int_0^\infty \lim_{t \to \infty} \langle \Delta D^+(\tau + t) \cdot \Delta D^-(t) \rangle e^{-i\omega \tau} d\tau, \tag{3}$$

where Re denotes the real part, and $\Delta D^\pm(t) = \Delta D^\pm(t) - \langle \Delta D^\pm(\infty) \rangle$ is the deviation of the dipole polarization operator $D^\pm(t)$ from its mean steady state value, and

$$D^+(t) = \mu_{01} a_1^\dagger a_2, \tag{4a}$$

$$D^-(t) = [D^+(t)]^\dagger, \tag{4b}$$

Then we rewrite the Eq. (2) in the form

$$\frac{d}{dt}\Psi = \mathbf{L}\Psi + \mathbf{I}, \quad (5)$$

where $\Psi = (\rho_{01}, \rho_{02}, \rho_{03}, \rho_{10}, \rho_{11}, \rho_{12}, \rho_{13}, \rho_{20}, \rho_{21}, \rho_{22}, \rho_{23}, \rho_{30}, \rho_{31}, \rho_{32}, \rho_{33})^T$, and $\mathbf{L}$ is a $15 \times 15$ matrix. The elements of $\mathbf{L}$ and $\mathbf{I}$ can be found explicitly from Eq. (2).

By means of the quantum regression theorem [32,33], the steady spectrum of the fluorescence spectrum from the state $|1\rangle$ to the ground state $|0\rangle$ can be obtained

$$S(\delta_k) = \text{Re}\{M_{11}\bar{\rho}_{11} + M_{12}\bar{\rho}_{12} + M_{13}\bar{\rho}_{13} + \sum_l N_{1l} I_l \bar{\rho}_{10}\}, \quad (6)$$

where

$$M_{ij} = [(z-L)^{-1}\big|_{z=i\delta_k}]_{i,j}, \quad N_{ij} = [L^{-1}(z-L)^{-1}\big|_{z=i\delta_k}]_{i,j}. \quad (7)$$

$\bar{\rho}_{ij}$ is the steady state population $(i = j)$ and atomic coherence $(i \neq j)$, which can be obtained by setting $\dot{\rho}_{ij} = 0$ and solving numerically Eq. (2). $\delta_k$ is the detuing between the fluorescence and the transition $|1\rangle \to |0\rangle$. In the next part, we will calculate the fluorescence spectrum by tuning the tunneling coupling, which depends on the barrier characteristics and the external electric field.

### III. RESULTS AND DISCUSSIONS

For simplicity, in this paper we mainly consider the case of resonant tunneling coupling. First, when none of the tunneling couplings is applied, the electron can not tunnel from QD 1 to the other QDs. Thus the system is like a single QD with one exited state $|1\rangle$ and one ground state $|0\rangle$, and the Mollow-type resonant fluorescence spectrum is obtained, as shown in Fig. 2(a). Such results have been achieved in earlier experiment in Ref. [19]. Then we fix $T_2 = 0$ and tune $T_1$ from 0 to 0.5, thus the electron can tunnel from QD 1 to QD 2, which is similar to a DQD system. In this case the quenching of the fluorescence emission is obtained [Fig. 2(b)]. When the detuning of the coupling laser is nonresonant, the fluorescence emission can take place again and acquire linewidth narrowing of the central peak, as shown in Fig. 2(c) and 2(d). As can be seen from the figure, with the smaller value of the detuning, we can obtain the narrower linewidth of the central peak.

In order to interpret the above results, we investigate the properties of the dressed levels under the coupling of the laser field $\Omega_c$ and the tunneling $T_1$. The energy eigenvalues of these dressed states [34] can be written as follows:

$$|\Psi_+\rangle = \sin\theta\sin\phi|0\rangle + \cos\phi|1\rangle + \cos\theta\sin\phi|2\rangle, \tag{8a}$$

$$|\Psi_0\rangle = \cos\theta|0\rangle - \sin\theta|2\rangle, \tag{8b}$$

$$|\Psi_-\rangle = \sin\theta\cos\phi|0\rangle - \sin\phi|1\rangle + \cos\theta\cos\phi|2\rangle. \tag{8c}$$

where

$$\tan 2\phi = \frac{\sqrt{4(\Omega_c^2 + T_1^2)}}{\delta_1}, \tag{9a}$$

$$\tan\theta = \frac{\Omega_c}{T_1}, \tag{9b}$$

And the eigenvalues of $|\Psi_i\rangle$ $(i = +, 0, -)$ is

$$\lambda_0 = 0, \tag{10a}$$

$$\lambda_\pm = \frac{\left(\delta_1 \pm \sqrt{\delta_1^2 + 4\left(\Omega_c^2 + T_1^2\right)}\right)}{2}. \tag{10b}$$

So we can see that both the state $|1\rangle$ and state $|0\rangle$ [see Fig. 1(c)] are split into three dressed levels [see Fig. 3(a)], which are $|\Psi_+\rangle$, $|\Psi_0\rangle$ and $|\Psi_-\rangle$ for the bare-state level $|1\rangle$, while $|\Psi_+'\rangle$, $|\Psi_0'\rangle$ and $|\Psi_-'\rangle$ for the bare-state level $|0\rangle$.

Therefore the fluorescence emission from the state $|1\rangle$ to the ground state $|0\rangle$ has nine dipole transitions in the dressed state representation. Note that though $|\Psi_i\rangle$ and $|\Psi_i'\rangle$ are different in constant energy by the energy difference between level $|1\rangle$ and $|0\rangle$, the dressed levels $|\Psi_i\rangle$ has the same expressions and the eigenvalues as the dressed levels $|\Psi_i'\rangle$ with $i = +, 0, -$.

Thus we can explain the quenching and narrowing of the fluorescence spectrum obtained in Fig. 2. First we plot in Fig. 3(b) the steady state population of the dressed date $|\Psi_i\rangle$ as a function of the coupling laser detuning $\delta_1$. When $\delta_1 = 0$, all the population is in the dressed state $|\Psi_0\rangle$. So the possible transition is $|\Psi_0\rangle \rightarrow |\Psi_+'\rangle$, $|\Psi_0\rangle \rightarrow |\Psi_0'\rangle$ and $|\Psi_0\rangle \rightarrow |\Psi_-'\rangle$. On the other hand, the transitions can occur only if dipole moment $|\langle \Psi_j' | \mathbf{P} | \Psi_i \rangle| \neq 0$, where $\mathbf{P} = \mathbf{\mu}_{01}|0\rangle\langle 1| + \mathbf{\mu}_{02}|0\rangle\langle 2|$ is the transition dipole moment operator of the bare state basis. And in the case of $\Gamma_{10} \gg \Gamma_{20}$, $\mathbf{P} = \mathbf{\mu}_{01}|0\rangle\langle 1|$. And from Eq. (8), the bare state $|1\rangle$ has no contribution to the dressed state $|\Psi_0\rangle$, therefore $|\langle \Psi_j' | \mathbf{P} | \Psi_i \rangle| = 0$ for the above three possible transitions and none of the transition can occur. As a result the quenching of the fluorescence emission is obtained.

Once the population is in state $|\Psi_0\rangle$, they will be trapped there, that is coherent population trapping (CPT). And the condition of $\delta_1 = \delta_2 = 0$ is the CPT condition. As the state $|\Psi_0\rangle$ is completely decoupled from the light field, we define the state $|\Psi_0\rangle$ as the dark state. So the quenching of the fluorescence emission is the result of the dark state, which is induced by the tunneling coupling between the dots. And unlike the dark state obtained in atomic system, in QD system we can use tunneling to achieve the necessary coherence, requiring no additional laser field.

And when $\delta_1 \neq 0$ and CPT condition is not satisfied, from Fig. 3(b) all the dressed state $|\Psi_i\rangle$ are populated and the dark state is destroyed. As a result the fluorescence emission can take place again [Fig. 2(c) and 2(d)].

It is well known that the widths of the fluorescence peaks are determined by the decay rate of the transition between the dressed levels $|\Psi_i\rangle$ to $|\Psi_j'\rangle$, which is proportional to the squared dipole moments $|\langle j'|\mathbf{P}|i'\rangle|$. And the central peak in Fig. 2(c) and 2(d)] arises from the transition of the between identical dressed levels of adjacent manifolds, which are $|\Psi_i\rangle \to |\Psi_i'\rangle$ $(i = +, 0, -)$. (The corresponding transitions are shown in Fig. 3(a) with the red dashed line.) We plot the decay rate $R_{ii}$ as a function of $T_1$ in Fig. 3(c) to explain the linewidths of the central peak under the nonresonant coupling laser. When $T_1$ is small ($T_1 = 0.2$), the value of $R_{00}$ is close to zero, which gives rise to a very sharp feature at line center. And when $T_1$ is increased to $T_1 = 0.5$, though the value of $R_{00}$ is increased a little bit, it is still small enough to give rise to the narrowing of the central peak. And in wide range of $T_1$, the value of $R_{++}$ and $R_{--}$ are much larger than those of $R_{00}$, thus the spectral feature at line center consists of a sharp peak superimposed on a broad profile. So we conclude that the narrowing of the spectrum is due to the slow decay of the dressed state, which arise from the tunneling coupling.

Next, we investigate fluorescence spectrum when both tunnelings are applied. In this case the electron can tunnel from QD 1 to QD 2 and from QD 2 to QD 3, which forms TQD system. With the resonant coupling of the laser field, when $T_2 = 0.2$, compared with Fig. 2(b) the fluorescence spectrum appears again, with one narrow central peak and two pairs of sidebands [Fig. 4(a)]. In the case of $T_2 = 0.5$, from Fig. 4(b) we can see that the intensity of the fluorescence is increased and the linewidth of the central peak is a little broadened. At the same time there is one pair of narrow inner sideband show up. And when the tunneling is increased to $T_2 = 1$, the intensity of the fluorescence emission continues to increase, as well as the linewidth of the central peak and the inner sideband. And also the location of the pair of inner narrow sideband gets farther away from the central peak [Fig. 4(c)]. When the detuning of the coupling field is tuned to $\delta_1 = T_2$, the quenching of the fluorescence emission is obtained again, as show in Fig. 4(d).

The above results can also be interpreted in the dressed state picture. Under the resonant coupling of the laser field and the two tunneling couplings, the energy eigenvalues of these dressed states can be written as follows:

$$\lambda_4 = -\lambda_1 = \kappa_+ , \tag{11a}$$

$$\lambda_3 = -\lambda_2 = \kappa_- , \tag{11b}$$

where

$$\kappa_\pm = \sqrt{\frac{(\Omega_c^2 + T_1^2 + T_2^2) \pm \sqrt{(\Omega_c^2 + T_1^2 + T_2^2)^2 - 4\Omega_c^2 T_2^2}}{2}} . \tag{12}$$

Then the dressed levels can be express as

$$|\Psi_1\rangle = C_{10}|0\rangle + C_{11}|1\rangle + C_{12}|2\rangle + C_{13}|3\rangle , \tag{13a}$$

$$|\Psi_2\rangle = -C_{20}|0\rangle + C_{21}|1\rangle - C_{22}|2\rangle + C_{23}|3\rangle , \tag{13b}$$

$$|\Psi_3\rangle = -C_{30}|0\rangle + C_{31}|1\rangle - C_{32}|2\rangle + C_{33}|3\rangle , \tag{13c}$$

$$|\Psi_4\rangle = C_{40}|0\rangle + C_{41}|1\rangle + C_{42}|2\rangle + C_{43}|3\rangle , \tag{13d}$$

where

$$C_{i0} = \frac{1}{D_i} \frac{\Omega_c \left( \lambda_i^2 - T_2^2 \right)}{T_1 T_2 \lambda_i} , \tag{14a}$$

$$C_{i1} = \frac{1}{D_i} \frac{\lambda_i^2 - T_2^2}{T_1 T_2} , \tag{14b}$$

$$C_{i2} = \frac{1}{D_i} \frac{\lambda_i}{T_2} , \tag{14c}$$

$$C_{i3} = \frac{1}{D_i} , \tag{14d}$$

$$D_i = \sqrt{1 + \left( \frac{\lambda_i}{T_2} \right)^2 + \left( \frac{\lambda_i^2 - T_2^2}{T_1 T_2} \right)^2 + \left( \frac{\Omega_c \left( \lambda_i^2 - T_2^2 \right)}{T_1 T_2 \lambda_i} \right)^2 } . \tag{14e}$$

And for simplicity, we rewrite Eq. (13) in the form

$$|\Psi_i\rangle = \sum C_{ik} |k\rangle, \ (i = 1,2,3,4; \ k = 0,1,2,3) \tag{15}$$

Similar to the DQD system, both state $|1\rangle$ and state $|0\rangle$ are split into four dressed levels [see Fig. 5(a)], which are $|\Psi_1\rangle$, $|\Psi_2\rangle$, $|\Psi_3\rangle$ and $|\Psi_4\rangle$ for the bare-state level $|1\rangle$, while $|\Psi_1'\rangle$, $|\Psi_2'\rangle$, $|\Psi_3'\rangle$ and $|\Psi_4'\rangle$ for the bare-state level $|0\rangle$. Therefore the fluorescence from the state $|1\rangle$ to the ground state $|0\rangle$ has sixteen dipole transitions in the dressed state representation. And also the dressed levels $|\Psi_i\rangle$ has the same expressions and the eigenvalues as the dressed levels $|\Psi_i'\rangle$ with

$i = 1, 2, 3, 4$.

First we explain the appearance of the fluorescence emission when tunneling $T_2$ is applied. In Fig. 5(b), with the resonant coupling of the laser field we plot the steady state population of the dressed level $|\Psi_i\rangle$ as a function of $T_2$. From the figure we can see that for all value of $T_2$, all dressed states are populated. On the other hand, all dipole moment $|\langle \Psi_j' | \mathbf{P} | \Psi_i \rangle| \neq 0$. So all transitions can occur and the electrons can not trapped in any of the dressed state. As a result the dark state is destroyed and fluorescence emission takes place.

Next we interpret the emerging of the narrow inner sideband [Fig. 4(b) and 4(c)]. This pair of peaks comes from the transition of $|\Psi_2\rangle \to |\Psi_3'\rangle$ and $|\Psi_3\rangle \to |\Psi_2'\rangle$ and the position of the peaks depend on the difference of the eigen energies $\lambda_2$ And $\lambda_3$. (The corresponding transitions are shown in Fig. 5(a) with the green dashed line.) We show in Fig. 5(c) the eigen energies $\lambda_i$ as a function of the tunneling $T_2$. From the figure we can see that the energies of the dressed levels $|\Psi_1\rangle$ and $|\Psi_4\rangle$ depend weakly on $T_2$, while the splitting between the dressed levels $|\Psi_2\rangle$ and $|\Psi_3\rangle$ increases with a larger value of $T_2$. As a result the position of the narrow inner peaks is becoming farther away from the central peak as $T_2$ is increased. So when $T_2$ is small, the inner sideband are so closed to the central peak that it can not be separated from the central peak [Fig. 4(a)]. While under the larger value of $T_2$, the narrow sideband gets farther away from the center and can be seen clearly [Fig. 4(b) and 4(c)].

Last we explain the narrowing of the central peak and the inner sideband obtained in Fig. 4. As we discussed above, the widths of the fluorescence peaks is proportional to the squared dipole moments $\left|\left\langle \Psi_j' \right| \mathbf{P} \left| \Psi_i \right\rangle\right|$. In TQD system, the transition dipole moment operator is $\mathbf{P} = \mu_{01}|0\rangle\langle 1| + \mu_{02}|0\rangle\langle 2| + \mu_{03}|0\rangle\langle 3|$. And in the case of $\Gamma_{10} \gg \Gamma_{20}(\Gamma_{30})$, $\mathbf{P} = \mu_{01}|0\rangle\langle 1|$. Thus the decay rate $R_{ij}$ can be calculated with the expression

$$R_{ij} = \left|C_{j0}\right|^2 \mu_{01}^{\ 2} \left|C_{i1}\right|^2 \quad (i, j = 1, 2, 3, 4). \tag{16}$$

The central peak comes from the transition $|\Psi_i\rangle \to |\Psi_i'\rangle$ $(i = 1\text{-}4)$, which are shown in Fig. 5(a) with the red dashed line. And we plot in Fig. 5(d) the decay rate $R_{ii}$ as a function of $T_2$ in order to explain the linewidths of the central peak. When $T_2$ is small, the value of $R_{22}(R_{33})$ is very small, which is the result of the narrowing of central peak. And when $T_2$ is increased, the value of $R_{22}(R_{33})$ does not changed much and still gives rise to the narrowing of the central peak. And in all values of $T_2$, the value of $R_{11}(R_{44})$ is much larger than those of $R_{22}(R_{33})$. Similar to the case of DQDs, the spectral feature at line center consists of a sharp peak superimposed on a broad profile. And the narrowing of the inner sideband, which arises from the transition $|\Psi_2\rangle \to |\Psi_3'\rangle$ and $|\Psi_3\rangle \to |\Psi_2'\rangle$, is also due to the slowly decay rate of the corresponding transition.

Alternatively, the TQD system can also be analyzed by diagonalizing the interaction with the tunneling $T_2$, and the corresponding dressed state is shown in Fig. 5(e). Under the coupling of $T_2$, the state $|2\rangle$ splits into two dressed levels $|2_\pm\rangle$ with splitting of $2T_2$ and creates two $\Lambda$ type subsystems. In both $\Lambda$ subsystems, the tunneling coupling $T_1$ is nonresonant with the transition $|1\rangle \to |2_+\rangle$ or $|1\rangle \to |2_-\rangle$, with detuning $\delta_2 = \pm T_2$. So with the resonant laser coupling, CPT condition is not satisfied in both $\Lambda$ systems, as a result the dark state is vanished and fluorescence emission takes place [Fig. 4(a)-4(c)]. As the strength of the tunneling $T_2$ is increased, the splitting between the level $|2_+\rangle$ and $|2_-\rangle$ becomes larger, thus the two $\Lambda$ systems are farther away from the CPT condition, so the intensity of the fluorescence is increased. Then we tune the detuning of the laser filed such that $\delta_1 = \pm T_2$, the dark state will be set up in one of the $\Lambda$ system and the fluorescence emission disappears [Fig. 4(d)].

At last we discuss the influence of the decay rate $\Gamma_{20}(\Gamma_{30})$ on fluorescence emission. The fluorescence comes from the spontaneous emission from dressed state $|\Psi_i\rangle$ to $|\Psi_j\rangle$, which can occur only if dipole moment $|\langle \Psi_j'|\mathbf{P}|\Psi_i\rangle| \neq 0$. For $\Gamma_{10} \gg \Gamma_{20}(\Gamma_{30})$, once the CPT condition is satisfied, all the dipole moments are zero and the quenching of the fluorescence emission can be obtained. Otherwise when $\Gamma_{20}(\Gamma_{30})$ can be compared with $\Gamma_{10}$, the transition dipole moment operator is $\mathbf{P} = \mu_{01}|0\rangle\langle 1| + \mu_{02}|0\rangle\langle 2| + \mu_{03}|0\rangle\langle 3|$, and $|\langle \Psi_j'|\mathbf{P}|\Psi_i\rangle| \neq 0$ even under the CPT condition. In this case the dark states is destroyed and the complete quenching of fluorescence emission can not be obtained. In Fig. 6 we display the fluorescence spectrum with various value of $\Gamma_{20}(\Gamma_{30})$. From the figure we can see that the bigger the value of $\Gamma_{20}(\Gamma_{30})$, the more population of the bare state $|1\rangle$, and the stronger the intensity of the fluorescence emission.

## IV. CONCLUSIONS

We have investigated the fluorescence spectrum from TQDs controlled by the tunneling and demonstrated that it is possible to use tunneling to induce dark states. With the tunneling coupling, interesting features such as quenching and narrowing of the fluorescence can be obtained. We have also interpreted the results in the dressed state basis of the two tunneling couplings and the laser field. The quenching of the fluorescence is due to the tunneling induced dark states, while the narrowing of the central peak is due to the slow decay rate of the dressed levels.

## ACKNOWLEDGMENTS

This work is supported by the financial support from the National Basic Research Program of China (Grant No. 2013CB933300), the National Natural Science Foundation of China (Grant No. 11304308, 61076064 and 61176046), and the Hundred Talents Program of Chinese Academy of Sciences.

**FIGURE CAPTIONS**

FIG. 1 (Color online) (a) The schematic of band structure without a gate voltage. (b) The schematic of band structure with a gate voltage. (c) The schematic of the level configuration of a TQD system.

FIG. 2 (Color online) Fluorescence spectrum $S(\omega)$ for $\Omega_c = 2$, $T_2 = 0$, $\delta_2 = \delta_3 = 0$, $\Gamma_{10} = 1$, $\Gamma_{20} = \Gamma_{30} = 10^{-6}\Gamma_{10}$, (a) $T_1 = 0$, $\delta_1 = 0$, (b) $T_1 = 0.5$, $\delta_1 = 0$, (c) $T_1 = 0.5$, $\delta_1 = 0.5$, (d) $T_1 = 0.2$, $\delta_1 = 0.2$.

FIG. 3 (Color online) (a) Dressed state under the coupling of the laser field $\Omega_c$ and the tunneling $T_1$. Properties of dressed levels as a function of the tunneling coupling $T_1$: (b) steady state populations of the dressed state $\rho_{\Psi_i}$ $(i = +, 0, -)$, (c) decay rates of the dressed state $R_{ii}$ $(i = +, 0, -)$. The parameters are the same as those in Fig. 2.

FIG. 4 (Color online) Fluorescence spectrum $S(\omega)$ for $\Omega_c = 2$, $T_1 = 0.5$, $\delta_2 = \delta_3 = 0$, $\Gamma_{10} = 1$, $\Gamma_{20} = \Gamma_{30} = 10^{-6}\Gamma_{10}$, (a) $T_2 = 0.2$, $\delta_1 = 0$, (b) $T_2 = 0.5$, $\delta_1 = 0$, (c) $T_2 = 1$, $\delta_1 = 0$, (d) $T_2 = 0.2$, $\delta_1 = 0.2$.

FIG. 5 (Color online) (a) Dressed state under the coupling of the laser field $\Omega_c$ and two tunneling $T_1$ and $T_2$. Properties of dressed levels as a function of the tunneling coupling $T_2$: (b) steady state populations of the dressed state $|\rho_{\Psi_i}\rangle$ $(i = 1-4)$, (c) the eigen energies $\lambda_{\Psi_i}$, (d) decay rates of the dressed state $R_{ii}$ $(i = 1-4)$. (e) Dressed state under the coupling the tunneling $T_2$. The parameters are the same as those in Fig. 4.

FIG. 6 (Color online) Fluorescence spectrum $S(\omega)$ for $\Omega_c = 2$, $T_1 = 0.5$, $T_2 = 0.5$, $\delta_1 = 0.5$, $\delta_2 = \delta_3 = 0$, $\Gamma_{10} = 1$, (a) $\Gamma_{20} = \Gamma_{30} = 10^{-6}\Gamma_{10}$, (b)

$\Gamma_{20} = \Gamma_{30} = 10^{-4}\Gamma_{10}$, (c) $\Gamma_{20} = \Gamma_{30} = 10^{-2}\Gamma_{10}$. (d) Steady state populations of bare state $|1\rangle$ as a function of $\Gamma_{20}$.

**FIGURE**

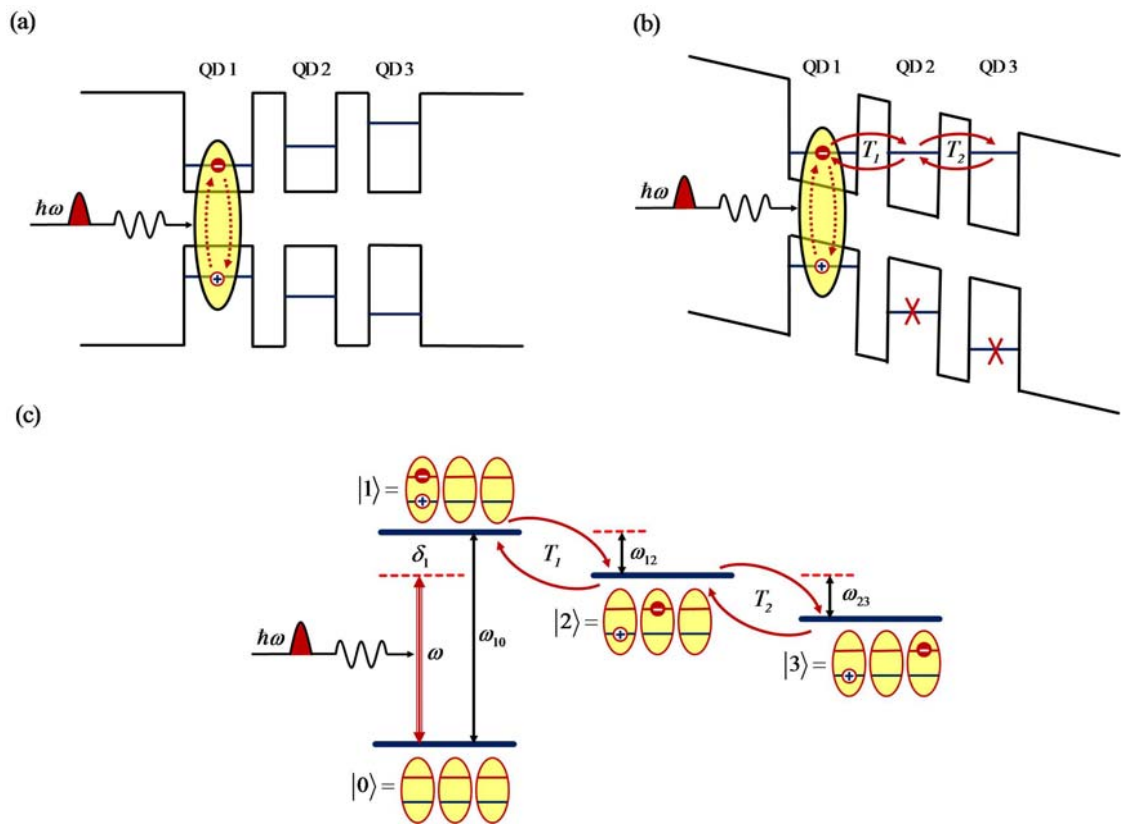

FIG. 1

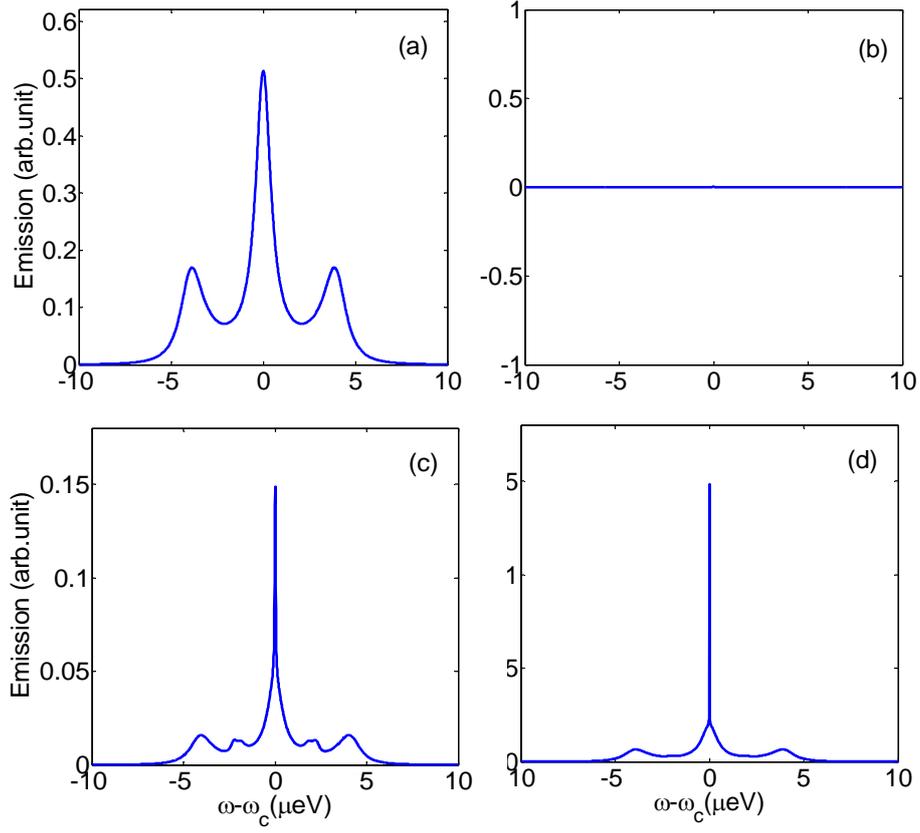

FIG. 2

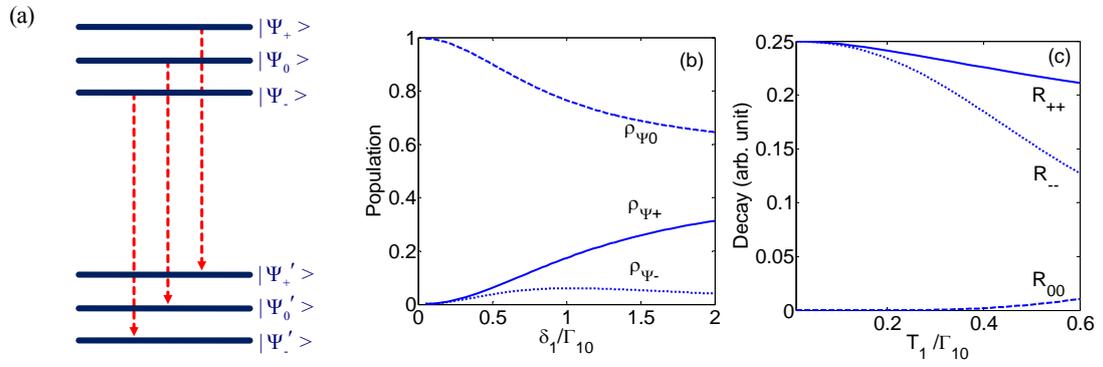

FIG. 3

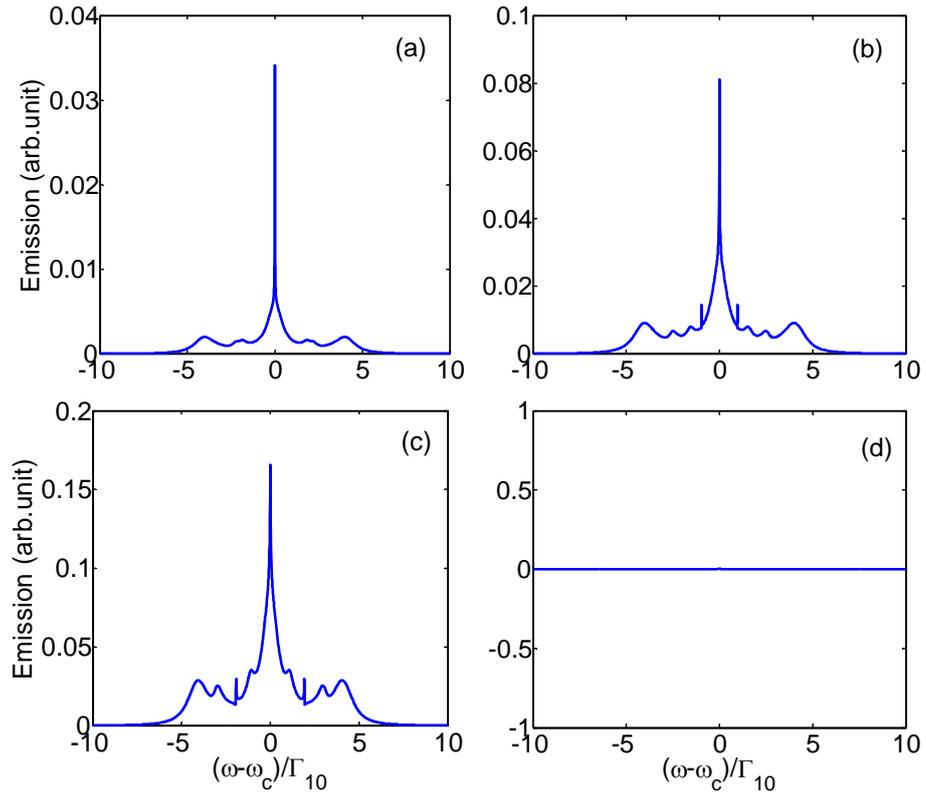

FIG. 4

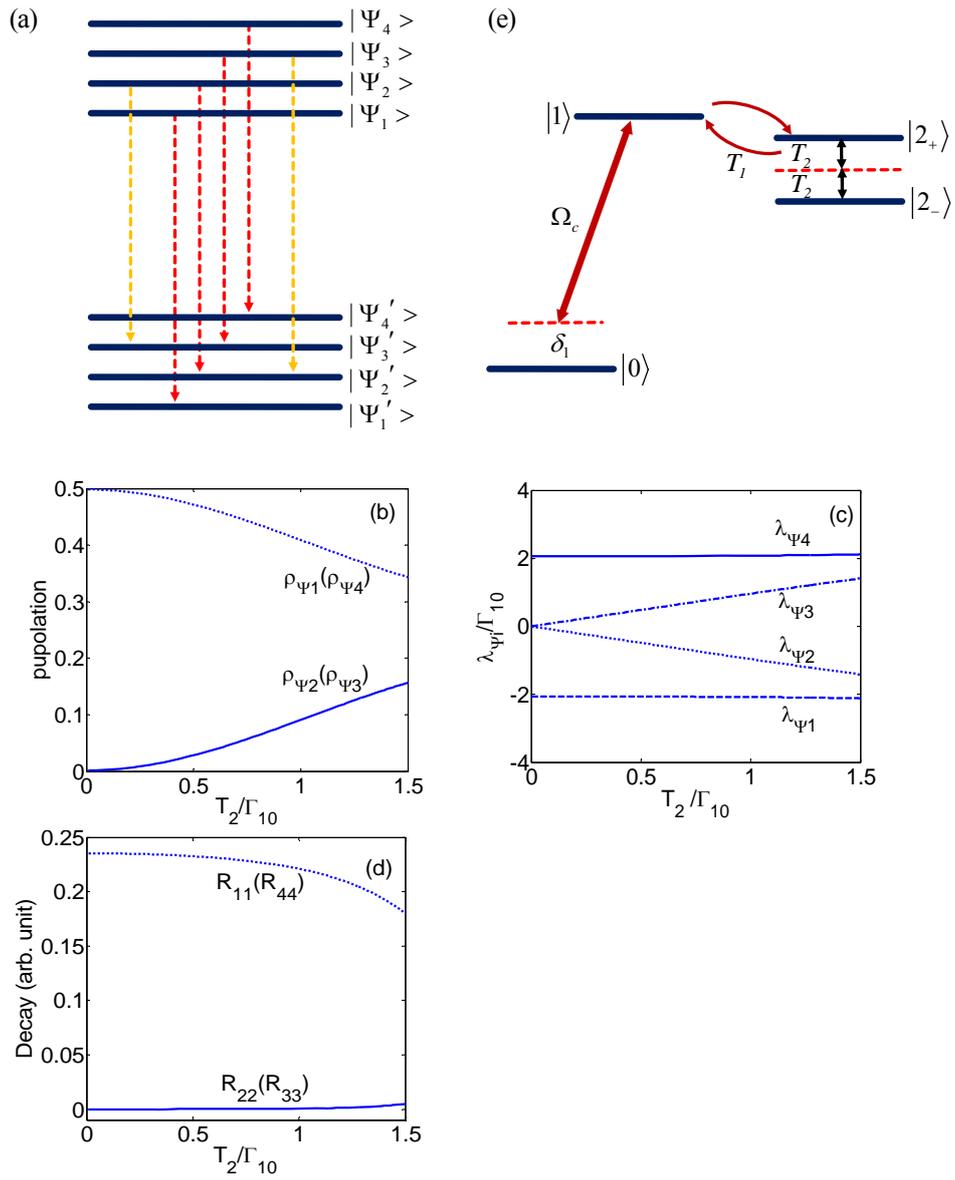

FIG. 5

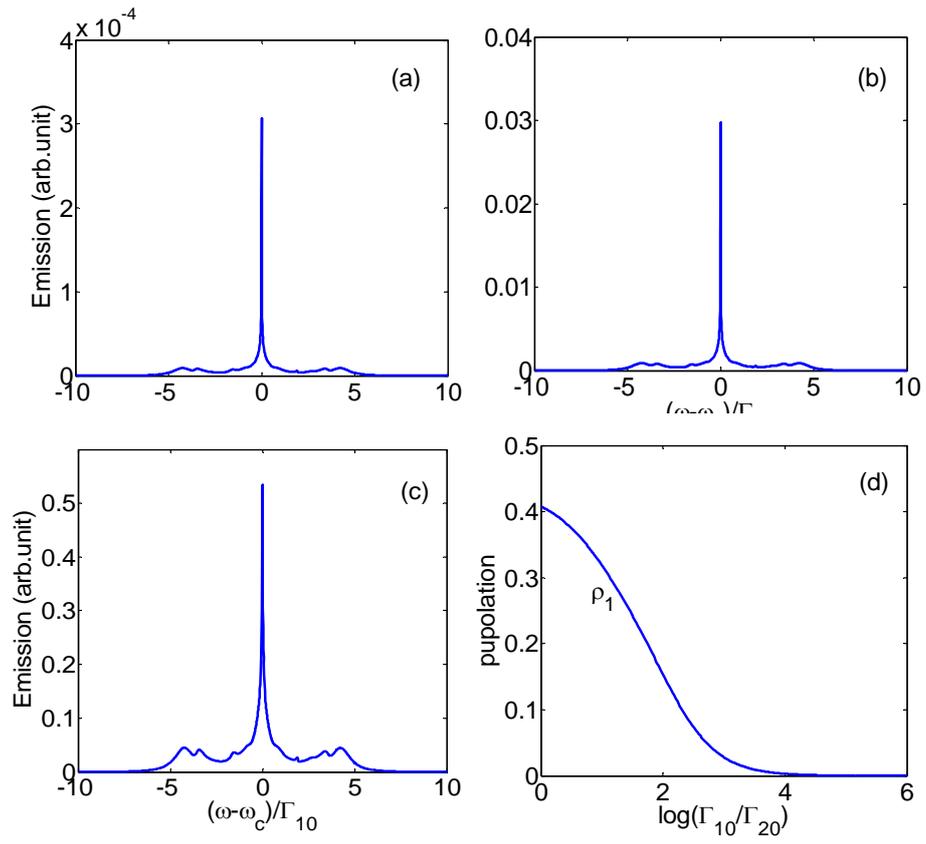

FIG. 6